\documentclass[journal=jaaucr, manuscript=article]{achemso}

\usepackage{comment}
\usepackage{xr}
\usepackage{siunitx} 
\usepackage[version=3]{mhchem} 
\usepackage{amsmath}
\bibstyle{achemso}

\makeatletter
\newcommand*{\addFileDependency}[1]{
  \typeout{(#1)}
  \@addtofilelist{#1}
  \IfFileExists{#1}{}{\typeout{No file #1.}}
}
\makeatother

\newcommand*{\myexternaldocument}[1]{%
    \externaldocument{#1}%
    \addFileDependency{#1.tex}%
    \addFileDependency{#1.aux}%
}

\myexternaldocument{SI}

\author{Lucy D. Whalley}
\email{l.whalley@northumbria.ac.uk}
\affiliation{Department of Mathematics, Physics and Electrical Engineering, Northumbria University, Newcastle upon Tyne, NE1 8QH, UK}

\author{Puck van Gerwen}
\affiliation{Department of Materials, Imperial College London, London SW7 2AZ, UK}

\author{Jarvist M. Frost}
\affiliation{Department of Physics, Imperial College London, London SW7 2AZ, UK}

\author{Sunghyun Kim}
\affiliation{Department of Materials, Imperial College London, London SW7 2AZ, UK}

\author{Samantha N. Hood}
\affiliation{Department of Materials, Imperial College London, London SW7 2AZ, UK}

\author{Aron Walsh}
\email{a.walsh@imperial.ac.uk}
\affiliation{Department of Materials, Imperial College London, London SW7 2AZ, UK}
\alsoaffiliation{Department of Materials Science and Engineering, Yonsei University, Seoul 03722, Korea}

\title[]
  {Giant Huang-Rhys Factor for Electron Capture by the Iodine Intersitial in Perovskite Solar Cells}

\begin{document}


\begin{tocentry}
\includegraphics{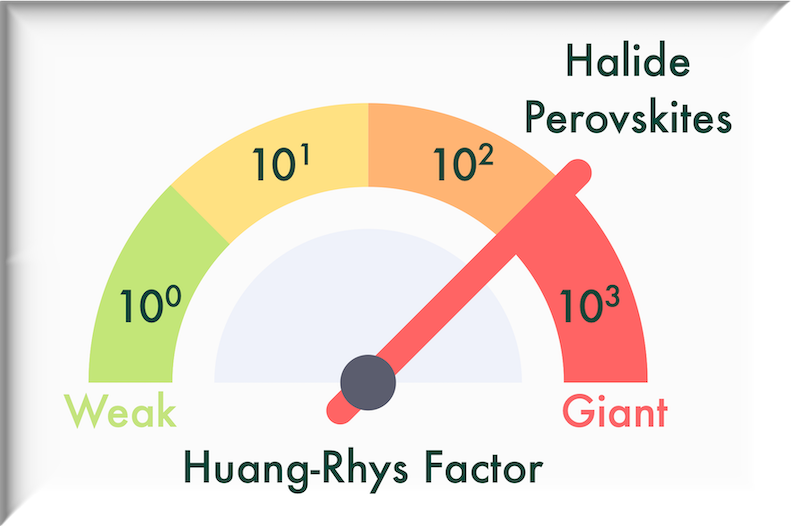} 
\end{tocentry}


\begin{abstract}
Improvement in the optoelectronic performance of halide perovskite semiconductors requires the identification and suppression of non-radiative carrier trapping processes. The iodine interstitial has been established as a deep level defect, and implicated as an active recombination centre. We analyse the quantum mechanics of carrier trapping. Fast and irreversible electron capture by the neutral iodine interstitial is found. The effective Huang-Rhys factor exceeds 300, indicative of the strong electron-phonon coupling that is possible in soft semiconductors. The accepting phonon mode has a frequency of \SI{53}{\per\centi\metre} and has an associated electron capture coefficient of \SI{1e-10}{\centi\metre\cubed\per\second}. The inverse participation ratio is used to quantify the localisation of phonon modes associated with the transition. We infer that suppression of octahedral rotations is an important factor to enhance defect tolerance.  
\end{abstract}

\section{Introduction}
The unusual defect chemistry and physics of lead halide perovskites has attracted significant attention.\cite{yin2014unusual,park2018point,motti2019defect}
Slow non-radiative electron-hole recombination is unusual for solution processed semiconductors, and supports high voltage and efficient light-to-electricity conversion in a solar cell.\cite{dequilettes2019charge}
While significant defect populations are expected based on equilibrium thermodynamics\cite{walsh2015self} of these soft crystalline materials, and solution processing introduces additional disorder,\cite{dunlap2018synthetic} the native defects don't appear to contribute to non-radiative recombination of electrons and holes. 
This behaviour of halide perovskites has been broadly termed ``defect tolerance".\cite{brandt2015identifying,walsh2017instilling,jaramillo2019praise}
Further improvement in the performance of halide perovskite devices requires suppression of non-radiative carrier capture and recombination events.\cite{luo2019minimizing}
In this report, we perform a quantum mechanical carrier-capture analysis of the interaction between electrons and the iodine interstitial in \ce{CH3NH3PbI3} (MAPI).

\begin{figure*}[]
\centering
  \includegraphics[width=0.6\textwidth]{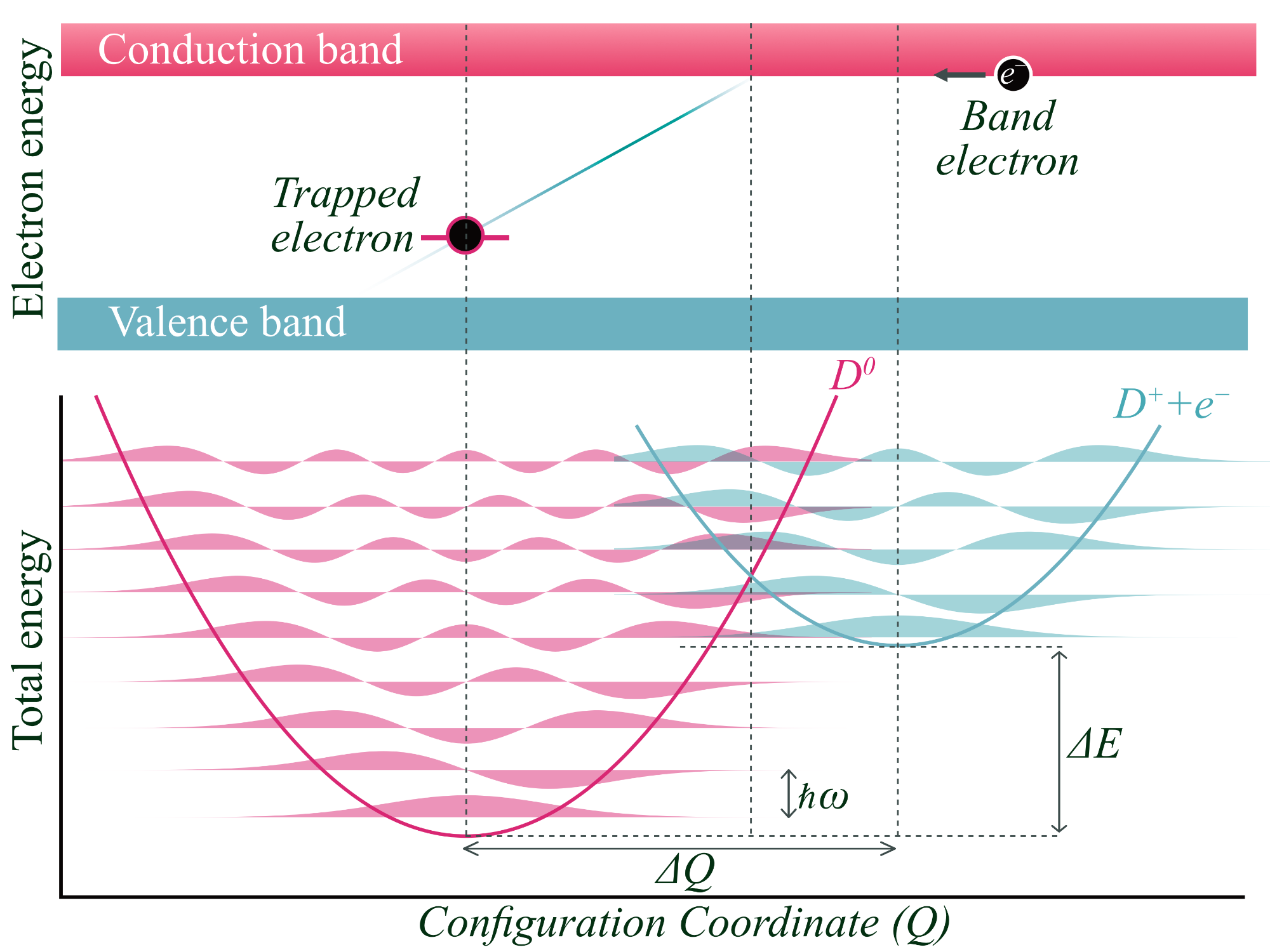}
  \caption{Electron capture by a charged defect in a semiconductor. 
  In the initial configuration (blue), there is a charged defect (D$^+$) and a free electron (e$^-$) residing in the conduction band.
  In the final configuration configuration (pink), the electron is captured to yield a neutral charge state (D$^0$). The coordinate $Q$ maps out the change in structure between the two configurations.}
\label{f1}
\end{figure*}

Shockley-Read-Hall (SRH) recombination is associated with the successive capture of an electron and hole following the photo-excitation (\ce{nil <=>[h\nu] e^- + h^+}) of a semiconductor.
The kinetics reduce to first-order in a heavily doped ($n$ or $p$ type) semiconductor; minority carrier capture becomes the rate-limiting process.
A necessary step is the change from delocalized-to-localized electronic wavefunctions, and multiple phonon ($\hbar\omega$) emission through the associated structural relaxation,\cite{Stoneham1981non} as illustrated in Figure \ref{f1}.
The excess electronic energy of the charge carriers is lost to heat. 
Taking the example of a neutral defect (D$^0$), the overall recombination process is
\begin{equation}
\ce{
D^0 <=>[h\nu]
D^0 + e^- + h^+ <=>[-\sum\hbar\omega] D^- + h^+ <=>[-\sum\hbar\omega] D^0}
\end{equation}

\subsection{Non-radiative carrier capture from first principles}
Making use of Fermi's golden rule, the carrier capture coefficient from an initial state \textit{i} to a final state \textit{f} can be described by
\begin{equation} 
\label{carriercapteqn}
    C = V \, \frac{2\pi}{\hbar} \, g \, W_\textrm{if}^2 \, \sum_m\Theta_m \, \sum_n |\langle\chi_{im}|\Delta Q|\chi_{fn}\rangle|^2  \times  \delta(\Delta E+m\hbar\omega_\textrm{i}-n\hbar\omega_\textrm{f})
\end{equation}
Here $V$ is the supercell volume, $g$ is the degeneracy of the final state, $W_\textrm{if}$ is the electron-phonon coupling matrix element, $\Theta_m$ is the thermal occupation of the vibration state $m$, $\langle\chi_{im}|\Delta Q|\chi_{fn}\rangle$ is the overlap of the vibrational wavefunctions $\chi$, and the Dirac $\delta(\Delta E+m\hbar\omega_\textrm{i}-n\hbar\omega_\textrm{f})$ ensures conservation of energy. 
Each of these quantities can be derived from density functional theory (DFT) calculations.\cite{alkauskas2014first,carriercapture} 

\begin{figure*}[t!]
\centering
  \includegraphics[width=\textwidth]{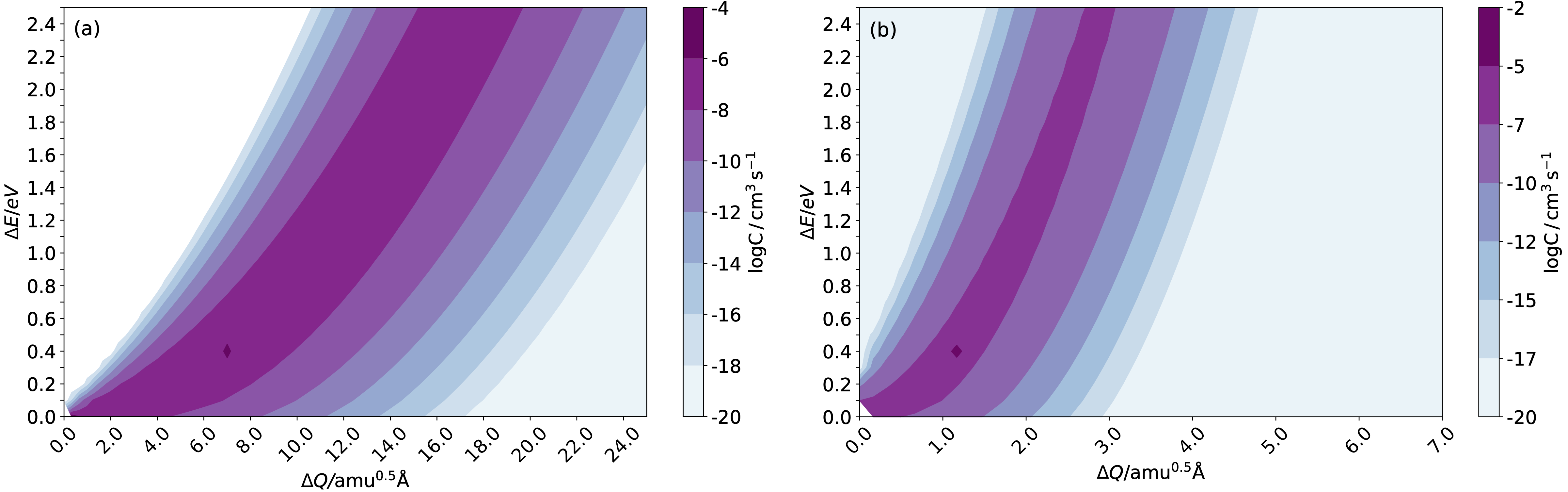}
  \caption[]{Variation of the carrier capture coefficient $C$ with the parameters $\Delta E$ and $\Delta Q$ using Equation \ref{carriercapteqn}.
This data is for the single frequency approximation and taking frequencies near the extrema of the physical range: (a) $\hbar\omega$ = 8 meV and (b) $\hbar\omega$ = 50 meV.}
\label{f2}
\end{figure*}

The meaning of Equation\ \ref{carriercapteqn} becomes clearer when you realise that $\Delta E$ and $\Delta Q$ refer to vertical and horizontal offsets in a configuration coordinate diagram, and that $\omega_\textrm{i}$ ($\omega_\textrm{f}$) is the frequency of the effective one-dimensional vibration in the initial (final) charge state (Figure \ref{f1}). 
Figure \ref{f2} provides a useful guide to the sensitivity of the underlying model parameters and the accessible range of $C$.

\section{Methods}

\subsection{Electronic structure}

A quantum mechanical treatment of electron capture was performed using the open-source \textsc{CarrierCapture} package.\cite{carriercapture}
The 1D Schr\"{o}dinger equation for the potential energy surface was solved using a finite difference method.
This builds on the approach of Alkauskas \textit{et al},\cite{alkauskas2014first} and the implementation has been applied to a range of semiconductors.\cite{kim2019anharmonic,kim2018identification}

The underlying electronic structures were calculated using density functional theory (DFT) as implemented in the GPU port of \textsc{VASP},\cite{Kresse1996a} using a plane wave basis set with an energy cut-off of \SI{400}{\electronvolt}.
Projection operators were optimised in real space with an accuracy of \SI{0.02}{\milli\electronvolt} per atom, and a $2\! \times\!2\!\times\! 2$ gamma centred Monkhorst-Pack mesh was used for the Brillouin zone integration.

So that no preference was given to a particular combination of octahedral tilts the starting point for atomic relaxation was MAPI in the pseudo-cubic phase.
The interstitial was placed in a 192-atom supercell built from an expansion of the 12-atom unit cell, using the transformation matrix $m_t$:
$$
m_t = \begin{bmatrix}
2 & -2 & 0 \\
2 & 2 & 0 \\
0 & 0 & 2 \\
\end{bmatrix}
$$
Ground states geometries were found using PBEsol functional,\cite{Perdew2008a} which has been shown to accurately describe the structures and phonons of these materials. 
The internal atomic coordinates were relaxed until the force acting on each atom was less than \SI{0.01}{\electronvolt\per\angstrom}. Defect formation energies were converged to within \SI{0.01}{\electronvolt} per formula unit between the 192- and 384-atom supercells.

The potential energy surface was calculated using the screened-exchange HSE06 functional\cite{Heyd2004a} ($\alpha$ = 0.43) including spin-orbit coupling (SOC), with total energy converged to within \SI{E-5}{\electronvolt}.
The electron-phonon coupling term is derived from wave functions calculated with the PBEsol functional and SOC.

\subsection{Lattice dynamics}

The harmonic phonon modes were calculated using a $2\! \times\!2\!\times\! 2$ supercell expansion (93 atoms including the iodine interstitial).
To evaluate the force-constant matrix the finite displacement method was used with displacements of \SI{0.01}{\angstrom}.
Forces were computed in VASP using a plane wave basis set of \SI{700}{\electronvolt}, a total energy convergence criterion of \SI{E-8}{\electronvolt} and the PBEsol functional.
A $2\! \times\!2\!\times\! 2$ gamma centred Monkhorst-Pack mesh was used for the Brillouin zone integration.
To extract the phonon eigenvectors and frequencies the \textsc{Phonopy} package\cite{Togo2015} was used.
The Inverse Participation Ratio was calculated using the \textsc{Julia-Phonons} package.\cite{juliaphonons}

\section{Results and Discussion}

\begin{figure*}
\centering
  \includegraphics[width=0.7\textwidth]{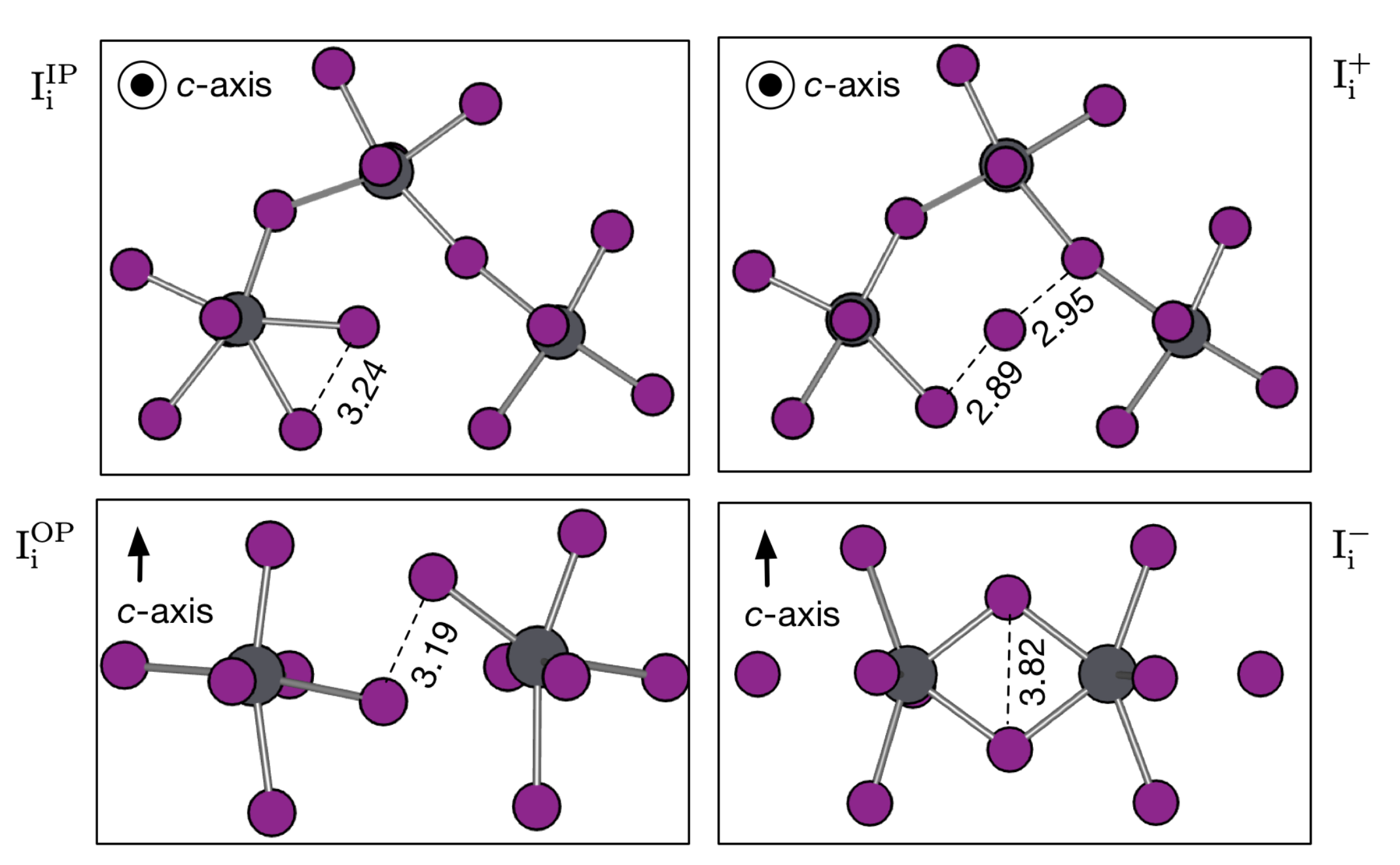}
  \caption[]{Defect geometries of $\mathrm{I}_\mathrm{i}^+$,$\mathrm{I}_\mathrm{i}^-$,$\mathrm{I}_\mathrm{i}^\mathrm{IP}$ and $\mathrm{I}_\mathrm{i}^\mathrm{OP}$ in \ce{CH3NH3PbI3}. IP indicates that the neutral defect is lying in the $ab$-plane. OP indicates that the defect is lying along the $c$-axis. All distances are measured in units of \AA. The iodine is coloured in purple and the lead in grey. For clarity, the organic cations are not show.}.
\label{f3}
\end{figure*}

Common vacancy defects do not introduce levels into the band gap of lead iodide perovskites, whilst iodine interstitials do.\cite{yin2014unusual} 
Interstitial iodine may be formed by the incorporation of excess iodine (\ce{I(g) <=> I_i}) 
or through Frenkel pair formation (\ce{I_I <=> V_I + I_i}).
There are three accessible charge states for the iodine interstitial (+/0/$-$).
The calculated geometries of the three charge states are shown in Figure \ref{f3}.

The neutral interstitial, rather than remaining as an isolated iodine atom in an interstitial region, bonds with a lattice iodine to form an I$_2^-$ complex. 
This is referred to as an H-center by the metal halide community and has a characteristic paramagnetic $S=\frac{1}{2}$ spin configuration.\cite{whalley2017h}
Iodine is well known to form polyiodide chains with bond lengths that are sensitive to the charge state.
The I--I bond length in solid orthorhombic crystalline iodine is \SI{2.67}{\angstrom}, which lengthens to \SI{3.23}{\angstrom} upon formation of $\mathrm{I}_2^-$.\cite{Chen1985}
Our bond lengths, calculated using a 193-atom pseudo-cubic supercell and plane wave basis set cut-off at \SI{400}{\electronvolt}, are within \SI{0.05}{\angstrom} of this value.
Owing to the octahedral tilting pattern, we distinguish between in-plane (IP) and out-of-plane (OP) configurations for this defect. 

For the positive charge state, an asymmetric trimer structure is found that is typical of $\mathrm{I}_3^-$. 
For example, the tri-iodide group in \ce{CsI_3} has interbond distances of \SI{2.82}{\angstrom} and \SI{3.10}{\angstrom}, with the longer bond possessing the majority of the additional charge.\cite{Finney1973}
In the negative charge state, the antibonding orbitals are filled, resulting in a split-interstitial configuration with the longest I--I bond length of all three charge states.

In the dark, the positive charge state is thermodynamically favoured in a p-type regime (E$_F$ close to the valence band) and the negative charge is favoured in an n-type regime (E$_F$ close to the conduction band).
The neutral interstitial is metastable but has been suggested to form under illumination,\cite{m2018iodine,Zhang2020iodine} which is our focus here.

Various first-principles studies have attempted to elucidate the nature of charge capture at the iodine interstitial site.
Fast non-radiative recombination \cite{Zhang2020iodine} and fast radiative recombination\cite{m2018iodine} at the neutral iodine interstitial have each been suggested.
This contradiction is despite both studies evaluating the electronic structure using the same Heyd-Scuseria-Ernserhof (HSE) hybrid density functional\cite{Heyd2004a} and incorporating spin-orbit coupling (SOC),
demonstrating the sensitivity of this system to the exact calculation parameters and defect geometries.
As an extension to static DFT, ab-initio molecular dynamics (MD) can be used to model the motion at room temperature. 
Large fluctuations in the halide vacancy defect energy level (up to \SI{1}{\electronvolt}),\cite{Cohen2019breakdown} and the formation of hole polarons that suppress charge recombination\cite{Wiktor2018mechanism} have both been reported.
However, a comprehensive MD analysis---using a hybrid functional and SOC--- remains computationally prohibitive.\cite{Cohen2019breakdown,Li2019influence}
We find a charge capture rate that is highly sensitive to the defect geometry; 
the importance of geometry relaxation, comparison with other computational reports, and the reference configuration for defect energetics are discussed in the Supplementary Information. 

\begin{figure*}
\centering
  \includegraphics[width=\textwidth]{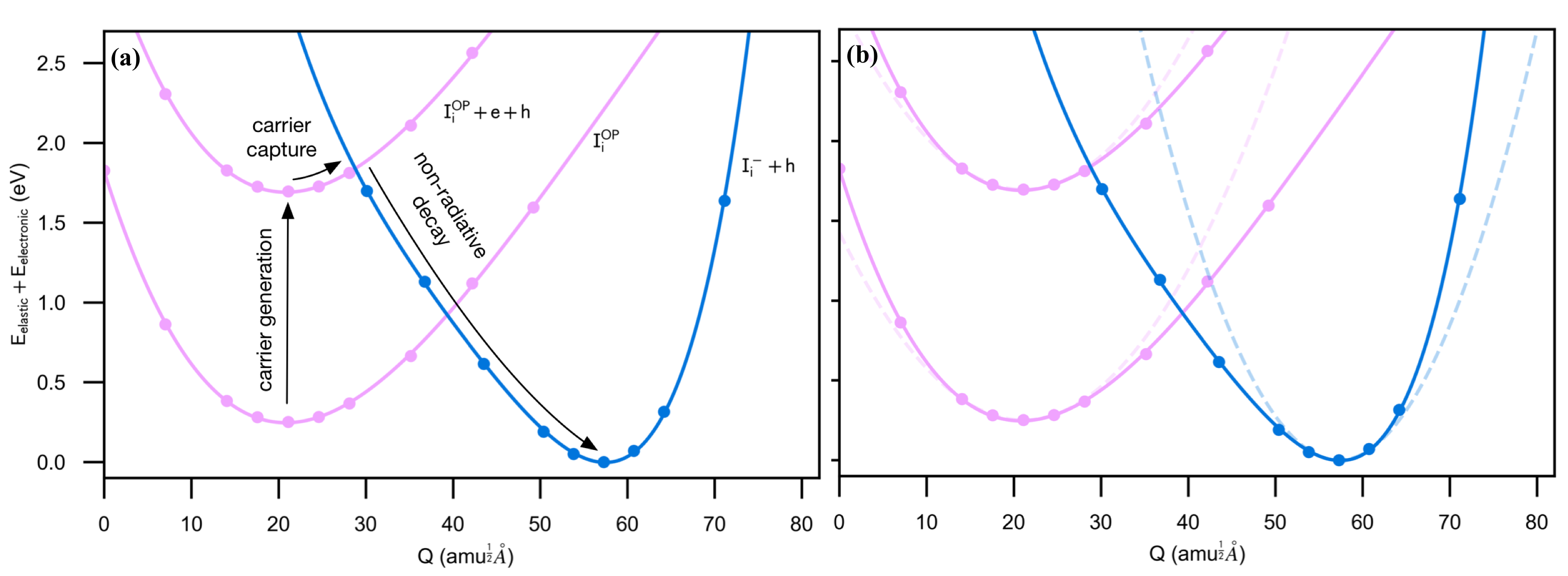}
  \caption[]{Configuration coordinate diagram for carrier capture by the iodine interstitial. 
  The DFT energies (solid circles) were calculated using the hybrid HSE06 functional. The coordinate $Q$, which corresponds to a linear combination of phonon modes that map between the two charge states, is defined in the main text. (a) To model electron trapping at the neutral iodine interstitial, the excited state of the system corresponds to the neutral defect with a photo-generated electron in the conduction band and hole in the valence band. The ground state corresponds to the negatively charged defect with a hole in the valence band. 
  (b) A comparison between the harmonic (dashed line) and anharmonic (solid line) PES. The curvatures are determined from second and fourth order spline fits to the DFT calculated energies.}
\label{f4}
\end{figure*}

We now consider the potential energy surface (PES) associated with sequential electron and hole capture, Figure \ref{f4}.
\begin{equation} 
\label{captureprocess}
\ce{
I_i <=>[h\nu] 
I_i + e^- + h^+ <=>[-\sum\hbar\omega]
I_i^- + h^+ <=>[-\sum\hbar\omega] I_i
}
\end{equation}
The coordinate $Q$ is defined as $\sqrt{\sum_i m_i \Delta r_i^2}$, where the sum is over all inorganic atoms $i$ with mass $m_i$ and a displacement from equilibrium of $\Delta r_i$.
After electron capture at \ce{I_i^{OP}}, not only the I dimer but the Pb atoms and the surrounding I atoms that form the octahedra relax significantly, as shown in Figure \ref{f3}. 
Analysis of the Pb-I-Pb angles in each charge state demonstrates that charge capture is associated with rotations of the inorganic octahedral cage (Figure S6).
As the large displacements of the heavy Pb and I atoms are involved, $\Delta Q$ is large ($\Delta Q=$ \SI{36}{amu\tothe{1/2}\angstrom}).
This value is double that typically found for non-radiative recombination centres in kesterites,\cite{kim2018identification} and demonstrates that there is strong coupling between the electronic charge state of the defect and the lattice distortion.

The definition of configuration coordinate $Q$ is not unique.
For example, we might define $\Delta Q$ in the MAPI:I$_i$ system as the root mean squared displacement of the two bonding iodine.
For this definition of $\Delta Q$ we find that electron trapping at the neutral iodine interstitial proceeds with a small geometrical rearrangement, $\Delta Q=$\SI{0.073}{\angstrom}, resulting in fast radiative electron capture. This small distortion of the iodine dimer has been reported previously.\cite{m2018iodine} However, this model excludes the large relaxation of the surrounding perovskite structure, leading to a significant underestimation of $\Delta Q$ and $S$. 

There are two classes of phonon mode associated with non-radiative transitions.
Promoting modes couple the initial and final electronic states by producing a sizeable electron-phonon coupling matrix element $W_\textrm{if}$.
Accepting modes take up the excess electronic energy after charge capture, resulting in a change of mean displacement.
Rather than consider each phonon mode in turn (which would be computationally prohibitive), we use the effective mode $Q$ to consider the accepting modes only.\cite{Stoneham1977non}

$W_\textrm{if}$, a pre-factor in Equation \ref{carriercapteqn}, is proportional to: 
i) the change in overlap between the initial (delocalised) and final (localised) single particle electron wavefunctions, as a function of $Q$; and ii) the change in energy between these two states.
For electron capture at the neutral iodine interstitial, $W_\textrm{if}$ is \SI{0.0036}{\electronvolt amu\tothe{-1/2}\angstrom\tothe{-1}}. 
This is comparable to an estimate for the maximum possible value, $W_\textrm{if}^\textrm{max}=$ \SI{0.0048}{\electronvolt amu\tothe{-1/2}\angstrom\tothe{-1}}, demonstrating that $Q$ has both accepting and promoting character, and justifying the use of the configuration coordinate. 
Further details of this comparison can be found in the Supplementary Information.

The Huang-Rhys factor, $S=\frac{\Delta E}{\hbar\omega}$, is the number of phonons emitted after carrier capture.
$S$ in the strong coupling regime ($S>> 1$) is typically associated with non-radiative carrier capture accompanied by multi-phonon emission.\cite{Stoneham1977non}
The harmonic PES of the iodine interstitial is soft, with an effective frequency of \SI{38}{\per\centi\metre} (\SI{4.7}{\milli\electronvolt}) and \SI{53}{\per\centi\metre} (\SI{6.6}{\milli\electronvolt}) for the neutral and negative states, respectively. The low frequency of the negative charge state, and large $\Delta E$, gives a `giant' Huang-Rhys factor of 350.
For comparison, a substitutional Si atom in GaAs (also known as the \textit{DX}-center), the archetypal defect exhibiting the large lattice relaxation,
has a Huang-Rhys factor of $S=75$ with $\omega= $\SI{81}{\per\centi\metre} and $\Delta Q= $\SI{9}{amu\tothe{1/2}\angstrom}.\cite{Lang:1979ix,kim2019anharmonic} 

To identify the normal modes that are associated with lattice relaxation we analyse the phonon dispersion of the negatively charged iodine interstitial. For each phonon eigenvector ($e_i$) at the gamma point of the Brillouin zone, we calculate the inverse participation ratio (IPR):
\begin{equation}
\mathrm{IPR} = \frac{\sum_{i=1}^N\left(|e_i|\right)^2}{\left(\sum_{i=1}^N|e_i|\right)^2},
\end{equation}
where $N$ is the number of phonon modes.\cite{Canisius_1985} A fully localised phonon mode has an IPR of 1. For a 96-atom supercell, a fully delocalised mode has an IPR of $\frac{1}{96}=0.0104$.
We find that the lowest energy resonant mode (with an $\textrm{IPR}=0.046$) has a frequency of \SI{53}{\per\centi\metre}, equal to the frequency of the negative charge state PES in the harmonic approximation. 
The agreement indicates that this resonant mode has strong accepting carrier and will be active in the uptake of excess electronic energy after the charge transition.
Additional analysis of the phonon modes associated with the defective crystal can be found in  Supplementary Information.

Finally, we consider the capture coefficients associated with sequential electron and hole capture, as outlined in Equation \ref{captureprocess}. The electron capture coefficient $C_n$ determines the rate of electron capture $R_n$ at a neutral iodine interstitial
\begin{equation}
    R_n=C_nN_\mathrm{t}n 
\end{equation}
where $N_\mathrm{t}$ is the density of neutral defect traps and $n$ is the electron density. 

It is evident from Figure \ref{f4} that an anharmonic description is necessary, as parabolic functions poorly describe the PES away from the equilibrium structures.
For capture processes where the atomic displacement is smaller, a harmonic oscillator model can be used to predict non-radiative capture coefficients using bulk material parameters (e.g. carrier effective mass, dielectric constant) and the defect energetics ($\Delta E$ and the charge transition level).\cite{Das2020what}
Unfortunately this model, whilst computationally economic and ideal for high-throughput type studies, is not valid for the iodine interstitial due to the anharmonicity of the PES.

In the harmonic picture, the iodine interstitial has an electron capture coefficient of \SI{8e-17}{\centi\metre\cubed\per\second}. 
Analysis of the vibrational wavefunctions for each PES shows that quantum tunnelling between the initial and final charge states is significant at energies below the classical barrier (Figure S5). 
Despite tunnelling, the capture coefficient is low due to the large electron capture barrier (\SI{600}{\milli\electronvolt}).
Fast radiative recombination, which is of the order \SI{1e-10}{\centi\metre\cubed\per\second} in MAPI,\cite{davies_bimolecular_2018} dominates in the harmonic approximation. 

In the more accurate anharmonic picture the electron capture coefficient is \SI{1e-10}{\centi\metre\cubed\per\second}, suggesting that non-radiative electron capture will compete with radiative capture. 
The electron capture barrier is \SI{148}{\milli\electronvolt}, much smaller than in the harmonic case, yielding a larger capture coefficient.
As in the harmonic picture, quantum tunnelling is significant (Figure S5).
Measuring the electron capture rate at the iodine interstitial is difficult as the trap density is hard to quantify and varies considerably with the material processing protocol. Nevertheless, a tentative approximation of the capture rate from a rate constant of \SI{2e7}{\per\second}\cite{Milot2015temperature} and assuming a trap density of \SI{1e16}{\per\centi\metre\cubed},\cite{Xing2014low-temperature} yields a capture coefficient of \SI{2e-9}{\centi\metre\cubed\per\second}. This estimate of the capture coefficient is within one order of magnitude of our calculated value for the anharmonic PES.

We observe a striking asymmetry between the electron capture barrier (\SI{148}{\milli\electronvolt}) and the hole capture barrier (\SI{924}{\milli\electronvolt}). 
This is due to the large lattice relaxation associated with iodine rearrangement upon electron capture. 
The asymmetry provides an explanation for the low trap-assisted recombination rate. 
Although electron capture at the neutral interstitial is fast, the subsequent capture of a hole at the negative iodine interstitial is energetically inaccessible for thermal electrons.

\section{Conclusions}

In conclusion, the soft nature of halide perovskites results in strong electron-phonon coupling and a large displacement of the surrounding inorganic octahedra following electron capture.
This relaxation process leads to a giant Huang-Rhys factor and facilitates fast non-radiative electron capture.

We expect to find similarly large, anharmonic lattice relaxations in other perovskites where dynamic octahedral tilting is evident (e.g. \ce{HC(NH2)2PbI3}, FAPI\cite{Weller2015cubic}) and, more generally, in mechanically soft semiconductors that are prone to structural disorder (e.g. metal-organic frameworks).
Defect tolerant perovskites (materials with low rates of non-radiative recombination) may be engineered by suppressing octahedral rotations in response to the changes in defect charge state. This might be achieved through elemental substitutions that result in local strain fields\cite{Yang2020} or through defect engineering.\cite{Ma2019}

On the basis of our results, the electron capture process at the neutral iodine interstitial in MAPI is irreversible; it is energetically unfavourable for the electron to be released into the band or annihilated by a hole.
An extension of this procedure to cover all native defects, while being computationally demanding, would help to understanding the nature of non-radiative losses in halide perovskites solar cells, as well as avenues to further enhance efficiency towards the radiative limit.

\begin{acknowledgement}

Calculations were performed on the Piz Daint supercomputer at the Swiss National Supercomputing Centre (CSCS) via the Partnership for Advanced Computing in Europe (PRACE) project pr51. 
Via our membership of the UK's HPC Materials Chemistry Consortium, which is funded by EPSRC (EP/L000202, EP/R029431), this work also used the ARCHER Supercomputing Service (http://www.archer.ac.uk). 
This work was supported by a National Research Foundation of Korea (NRF) grant funded by the Korean government (MSIT) (No. 2018R1C1B6008728) and the H2020 Programme under the project STARCELL (H2020‐NMBP‐03‐2016‐720907). J.M.F. is supported by a Royal Society University Research Fellowship (URF-R1-191292). 

\end{acknowledgement}

\begin{suppinfo}

Data access, calculation procedure for defect properties, additional analysis of the carrier capture rate.

\end{suppinfo}


\clearpage

\bibliography{lib}

\end{document}